# Spatial correlation between submillimetre and Lyman-α galaxies in the SSA 22 protocluster

Yoichi Tamura[1,2], Kotaro Kohno[3], Kouichiro Nakanishi[2,4], Bunyo Hatsukade[3], Daisuke Iono[3,4], Grant W. Wilson[5], Min S. Yun[5], Tadafumi Takata[2], Yuichi Matsuda[2], Tomoka Tosaki[4], Hajime Ezawa[4], Thushara A. Perera[5], Kimberly S. Scott[5], Jason E. Austermann[5], David H. Hughes[6], Itziar Aretxaga[6], Aeree Chung[5], Tai Oshima[4], Nobuyuki Yamaguchi[4], Kunihiko Tanaka[4] & Ryohei Kawabe[4]

[1]Department of Astronomy, University of Tokyo, Hongo, Bunkyo-ku, Tokyo 113-0033, Japan. [2]National Astronomical Observatory of Japan, 2-21-1 Osawa, Mitaka, Tokyo 181-8588, Japan. [3]Institute of Astronomy, University of Tokyo, 2-21-1 Osawa, Mitaka, Tokyo 181-0015, Japan. [4]Nobeyama Radio Observatory, National Astronomical Observatory of Japan, Minamimaki, Minamisaku, Nagano 384-1305, Japan. [5]Department of Astronomy, University of Massachusetts, Amherst, Massachusetts 01003, USA. [6]Instituto Nacional de Astrofísica, Optica y Electrónica, Aptdo. Postal 51 y 216, 72000 Puebla, Puebla, Mexico.

**Lyman-α emitters are thought to be young, low-mass galaxies with ages of ~$10^8$ yr (refs 1, 2). An overdensity of them in one region of the sky (the SSA 22 field) traces out a filamentary structure in the early Universe at a redshift of $z \approx 3.1$ (equivalent to 15 per cent of the age of the Universe) and is believed to mark a forming protocluster[3,4]. Galaxies that are bright at (sub)millimetre wavelengths are undergoing violent episodes of star formation[5–8], and there is evidence that they are preferentially associated with high-redshift radio galaxies[9], so the question of whether they are also associated with the most significant large-scale structure growing at high redshift (as outlined by Lyman-α emitters) naturally arises. Here we report an imaging survey of 1,100-μm emission in the SSA 22 region. We find an enhancement of submillimetre galaxies near the core of the protocluster, and a large-scale correlation between the submillimetre galaxies and the low-mass Lyman-α emitters, suggesting synchronous formation of the two very different types of star-forming galaxy within the same structure at high redshift. These results are in general agreement with our understanding of the formation of cosmic structure.**

Many different populations of young star-forming galaxies in the early Universe are known, but the relations among them and to the cosmic large-scale structure are still not well understood. The members of one of these populations are characterized by their strong Lyman-α (Lyα) emission (luminosity, $L_{Ly\alpha} \gtrsim 10^{42}$ erg s$^{-1}$), arising from ionized gas; their deficiency in ultraviolet continuum emission, which is interpreted as having a relatively small stellar component[1] ($M_{star} \lesssim 10^{9} M_{\odot}$, where $M_{\odot}$ is the solar mass); and their small size[2] ($\lesssim$ 1 kpc in diameter). The Lyα emitters towards SSA 22 trace a large-scale (~10 arcmin) filamentary structure that extends over several tens of megaparsecs (co-moving scale) and which may be the largest protocluster yet detected at high redshift[4].

Massive galaxies forming through accretion and mergers of small galaxies in such high-density environments are expected to be dust-obscured starbursts, which are too faint to detect at optical wavelengths but are observed as submillimetre-bright galaxies (SMGs). It is known from previous studies that SMGs have molecular gas reservoirs of $10^{10} M_{\odot}$–$10^{11} M_{\odot}$ (ref. 10) for their star-formation activities, suggesting that they are progenitors of massive elliptical galaxies seen in the cores of present-day clusters[8,11]. Individual ~5-arcmin$^2$-wide, deep submillimetre surveys in the direction of powerful, high-redshift radio galaxies, which are also believed to trace protoclusters[12], have presented tentative evidence for an enhancement in the number density of submillimetre sources around them[9]. Although these observations were limited in sensitivity and spatial coverage, they support the idea that SMGs are related to large-scale structure. To better understand the connection between the formation of massive galaxies and large-scale structure, we mapped the large-scale distribution of (sub)millimetre-bright, dusty starburst galaxies in the SSA 22 protocluster.

We carried out a wide-area (390-arcmin$^2$) survey of the SSA 22 field at 1,100 μm using the AzTEC camera[13] mounted on the Atacama Submillimeter Telescope Experiment (ASTE)[14], Chile (see also Supplementary Fig. 1). Our AzTEC map (Fig. 1a), which is more than 20 times larger than any of the existing maps at submillimetre wavelengths in this field (see, for example, refs 15–17), is wide enough to cover the region of the entire protocluster. We have detected 30 SMGs with signal-to-noise ratios $s/n \geq 3.5$ (a full source list is given in Supplementary Table 1). Their intrinsic flux densities are in the range 1.9–8.4 mJy (1 Jy = $10^{-23}$ erg s$^{-1}$ cm$^{-2}$ Hz$^{-1}$), corresponding to far-infrared luminosities of $L_{FIR} > 4 \times 10^{12} L_{\odot}$ (where $L_{\odot}$ is the solar luminosity) if we assume an emissivity index of $\beta = 1.5$, a dust temperature of $T_{dust} = 40$ K and that the sources are located at $z$ = 2–6. The inferred star-formation rates of the 1,100-μm sources are ~$10^3 M_\odot$ yr$^{-1}$, assuming that star formation is the dominant mechanism that heats the dust.

The most prominent new finding is that the distribution of the brighter (≥2.7 mJy) half of the 1,100-μm sources (15 of the 30, hereafter termed 'bright' SMGs; Table 1), which suffer little from incompleteness and false detections (Supplementary Figs 2 and 3), appears to be correlated with the high-density region of Lyα emitters[4], as seen in Fig. 1b. A concentration of bright SMGs ~5 arcmin northwest of the field centre is evident. Seven of the 15 bright SMGs (47%) are concentrated within a 50-arcmin$^2$ region in the direction that has a large-scale filamentary structure of Lyα emitters ~50 Mpc in depth (see fig. 1 of ref. 18). The number density over this region is 2–3 times higher than those found in blank-field surveys at 1,100 μm (ref. 19). Furthermore, the three most significant sources ($8.4^{+0.8}_{-1.0}$, $4.4^{+0.9}_{-0.8}$, $4.1^{+1.0}_{-0.8}$ mJy) are all located close (<4.5 arcmin) to the peak of the Lyα emitter overdensity. Photometric redshift estimates for the SMGs based on their radio and 24–1,100-μm flux ratios (Supplementary Fig. 4) indicate that they are probably at high redshift ($z$ > 1). The redshift estimates also suggest that some fraction of the bright SMGs, including the three most significant sources towards SSA 22, can be located at $z$ = 3.1 and may mark the local peak of the underlying mass distribution in the protocluster.

A two-point angular cross-correlation function is often used in determining the fractional increase in the probability of finding a source of one population within a unit solid angle as a function of angular distance from a source of another population, relative to a random distribution. We use an angular cross-correlation function to quantify the degree of cohabitation between the Lyα emitters and the bright SMGs. Figure 2 shows the cross-correlation function, which reveals strong correlation signals at angular distances less than 5 arcmin for the bright sample, suggesting close association of the Lyα emitters with the bright SMGs that are most likely embedded in the more massive dark haloes. Monte Carlo simulations (Supplementary Information) also show a correspondence between the two distributions, at a 97.3% significance level, further supporting the positional association of Lyα emitters with bright SMGs.

The gravitational lensing magnification of background galaxies by foreground large-scale structure would immediately preclude the physical connection between the galaxies and the foreground structure. Some authors[20,21] have reported correlations between bright (sub)millimetre sources and optically selected low-redshift galaxies (mostly at $z < 1$) in other regions of the sky. In general, SMGs are often found at high redshift (median, $z = 2.2$; ref. 22), and the maximal gravitational lensing magnification for a background galaxy at $z \gtrsim 2$ occurs when the foreground lensing structure is at $z \approx 0.5$. Therefore, they concluded that the correlation signal is most likely the result of amplification of background SMGs due to gravitational weak lensing by the foreground low-redshift galaxies. By contrast, the origin of the correlation signals in SSA 22 is most likely intrinsic to the large-scale structure in which both populations, SMGs and Lyα emitters, are embedded. Because the redshift estimates for the SMGs place them at distances coeval with the Lyα emitters, it is unlikely that the correlation seen in SSA 22 is due to amplification of a much higher-redshift ($z \gg 3.1$) SMG population lensed by the structure traced by the Lyα emitters, which are all located at $z = 3.1$ (not $z \approx 0.5$).

We do not detect the dust emission from individual Lyα emitters at the sensitivity of our 1,100-μm observations. This is a strong indication that SMGs and Lyα emitters are different populations, even though the Lyα emitters are spatially correlated with the SMGs. Of the 166 Lyα emitters within our 1,100-μm coverage, none are within the $2\sigma$ error circle (~26-arcsec diameter for $3.5 < s/n < 4.5$ and $\lesssim 20$ arcsec for $s/n > 4.5$) of an SMG; on average, we expect 2–3 SMGs to have a chance to be associated with a Lyα emitter in AzTEC's 28-arcsec beam if 30 SMGs and 166 Lyα emitters are randomly scattered in the 390-arcmin$^2$ region of our survey. To estimate the dust mass of a typical Lyα emitter in SSA 22, we stack the 1,100-μm images on the positions of the 166 Lyα emitters. We see no dust emission above 107 μJy ($2\sigma$) at 1,100 μm, and derive limits on far-infrared luminosity of $L_{\mathrm{FIR}} < 1.9 \times 10^{11} L_{\odot}$ and $L_{\mathrm{FIR}} < 1.7 \times 10^{12} L_{\odot}$ for $\beta = 1.5$ and, respectively, $T_{\mathrm{dust}} = 40$ K and $T_{\mathrm{dust}} = 70$ K. These luminosities correspond to respective dust masses of $M_{\mathrm{dust}} < 1.4 \times 10^{7} M_{\odot}$ and $M_{\mathrm{dust}} < 5.8 \times 10^{6} M_{\odot}$, assuming a dust emissivity of $\kappa_{850\,\mu\mathrm{m}} = 0.15$ m$^2$ kg$^{-1}$ (ref. 23). This limit is 3–40 times lower than the dust masses previously derived[24,25] for Lyα emitters at $z = 6.5$. Of course, the result from a simple stacking analysis cannot strongly constrain the dust properties of the Lyα emitter population. Nevertheless, this limit is 1–2 orders of magnitude lower than the average dust mass found in the population of SMGs, supporting the argument that Lyα emitters are on average less dust obscured[1] than SMGs.

These results provide evidence in favour of the synchronous formation of two very different types of high-redshift star-forming galaxy, SMGs and Lyα emitters, within the same cosmic structure. Although the formation process of SMGs is not yet fully understood, the observational evidence shown here suggests that they may form preferentially in regions of high mass concentration, which is consistent with predictions from the standard model of hierarchical structure formation[26,27]: we are presumably observing a galaxy-formation site where large-scale accumulation of baryonic matter is occurring within the large dark matter halo. Millimetre/submillimetre interferometric identifications followed by accurate measurements of the SMG redshifts will allow us to investigate this further.

**Supplementary Information** is linked to the online version of the paper at www.nature.com/nature.

**Acknowledgements** We acknowledge T. Yamada and T. Hayashino for providing the Lyα emitter catalogue. We are grateful to H. Hirashita, T. Suwa, T. Kodama, M. Sameshima, M. Hayashi, T. T.

Takeuchi and S. Komugi for discussions. We thank M. Uehara and the ASTE and AzTEC staff for their support. The ASTE project is lead by Nobeyama Radio Observatory, in collaboration with the University of Chile, the University of Tokyo, Nagoya University, Osaka Prefecture University, Ibaraki University, and Hokkaido University. This work is based in part on archival data obtained with the NASA Spitzer Space Telescope.

**Author Contributions** K.N., Y.T., T. Takata, K.K. and R.K. designed and proposed the survey. Y.T., K.K., K.N., B.H., D.I. and T. Tosaki conducted the observing runs for two months. G.W.W., T.A.P., J.E.A. and K.S.S. developed the AzTEC instrument and the fundamental AzTEC reduction pipeline. H.E., D.H.H., I.A, T.O., N.Y. and K.T. contributed to the operation of AzTEC and ASTE during the survey. Y.T. and B.H. processed the raw AzTEC data, carried out simulations to create a source catalogue and computed the correlation functions. M.S.Y. and A.C. processed the Very Large Array 20-cm data. Y.M. provided the Lyα emitter catalogue and contributed to discussions, especially on Lyα emitters. All the authors discussed the results.

**Author Information** Reprints and permissions information is available at www.nature.com/reprints. Correspondence and requests for materials should be addressed to Y.T. (yoichi.tamura@nao.ac.jp).

**Table 1 The bright SMG sample found in SSA 22**

| Source name | Coordinate (J2000) | | Flux density (mJy) | | s/n |
|---|---|---|---|---|---|
| | RA (h:min:s) | Dec. | $S_{\rm observed}$* | $S_{\rm deboost}$† | |
| SSA22-AzTEC1 | 22:17:32.42 | +0° 17′ 35.5″ | 8.7 ± 0.7 | $8.4^{+0.8}_{-1.0}$ | 12.8 |
| SSA22-AzTEC2 | 22:17:42.38 | +0° 16′ 59.3″ | 4.9 ± 0.7 | $4.4^{+0.9}_{-0.8}$ | 7.2 |
| SSA22-AzTEC3 | 22:17:18.85 | +0° 18′ 0.0″ | 4.7 ± 0.7 | $4.1^{+1.0}_{-0.8}$ | 6.8 |
| SSA22-AzTEC4 | 22:18:14.37 | +0° 9′ 53.1″ | 5.6 ± 0.9 | $4.7^{+1.2}_{-1.0}$ | 6.2 |
| SSA22-AzTEC5 | 22:17:10.77 | +0° 14′ 11.8″ | 4.0 ± 0.7 | $3.3^{+1.0}_{-0.8}$ | 5.6 |
| SSA22-AzTEC6 | 22:17:20.07 | +0° 20′ 11.0″ | 4.0 ± 0.7 | $3.3^{+1.0}_{-0.8}$ | 5.6 |
| SSA22-AzTEC7 | 22:17:40.82 | +0° 12′ 47.6″ | 3.6 ± 0.7 | $3.1^{+0.8}_{-0.9}$ | 5.2 |
| SSA22-AzTEC8 | 22:18:5.65 | +0° 6′ 42.0″ | 4.9 ± 1.0 | $3.9^{+1.2}_{-1.2}$ | 5.0 |
| SSA22-AzTEC9 | 22:17:54.40 | +0° 19′ 29.5″ | 3.6 ± 0.7 | $2.9^{+0.9}_{-0.9}$ | 5.0 |
| SSA22-AzTEC10 | 22:17:34.03 | +0° 13′ 46.8″ | 3.4 ± 0.7 | $2.8^{+0.9}_{-0.9}$ | 4.8 |
| SSA22-AzTEC11 | 22:17:29.64 | +0° 20′ 24.4″ | 3.3 ± 0.7 | $2.7^{+0.9}_{-0.9}$ | 4.7 |
| SSA22-AzTEC12 | 22:17:36.04 | +0° 4′ 0.2″ | 4.0 ± 0.9 | $3.1^{+1.1}_{-1.1}$ | 4.5 |
| SSA22-AzTEC13 | 22:18:5.95 | +0° 11′ 41.9″ | 3.3 ± 0.7 | $2.7^{+0.9}_{-1.0}$ | 4.5 |
| SSA22-AzTEC14 | 22:17:0.34 | +0° 10′ 42.6″ | 3.7 ± 0.9 | $2.7^{+1.2}_{-1.2}$ | 4.2 |
| SSA22-AzTEC15 | 22:16:57.60 | +0° 19′ 22.8″ | 4.1 ± 1.0 | $2.9^{+1.4}_{-1.3}$ | 4.2 |

A full list of the 30 submillimetre galaxies is given in Supplementary Table 1. Note that the astrometric accuracy of the catalogue is ≈10 arcsec.
*Observed flux density before flux bias correction, plus the 1$\sigma$ error.
†Deboosted flux density (flux density corrected for the flux bias due to confusion noise using the method described elsewhere[28]), plus the 68% confidence interval.

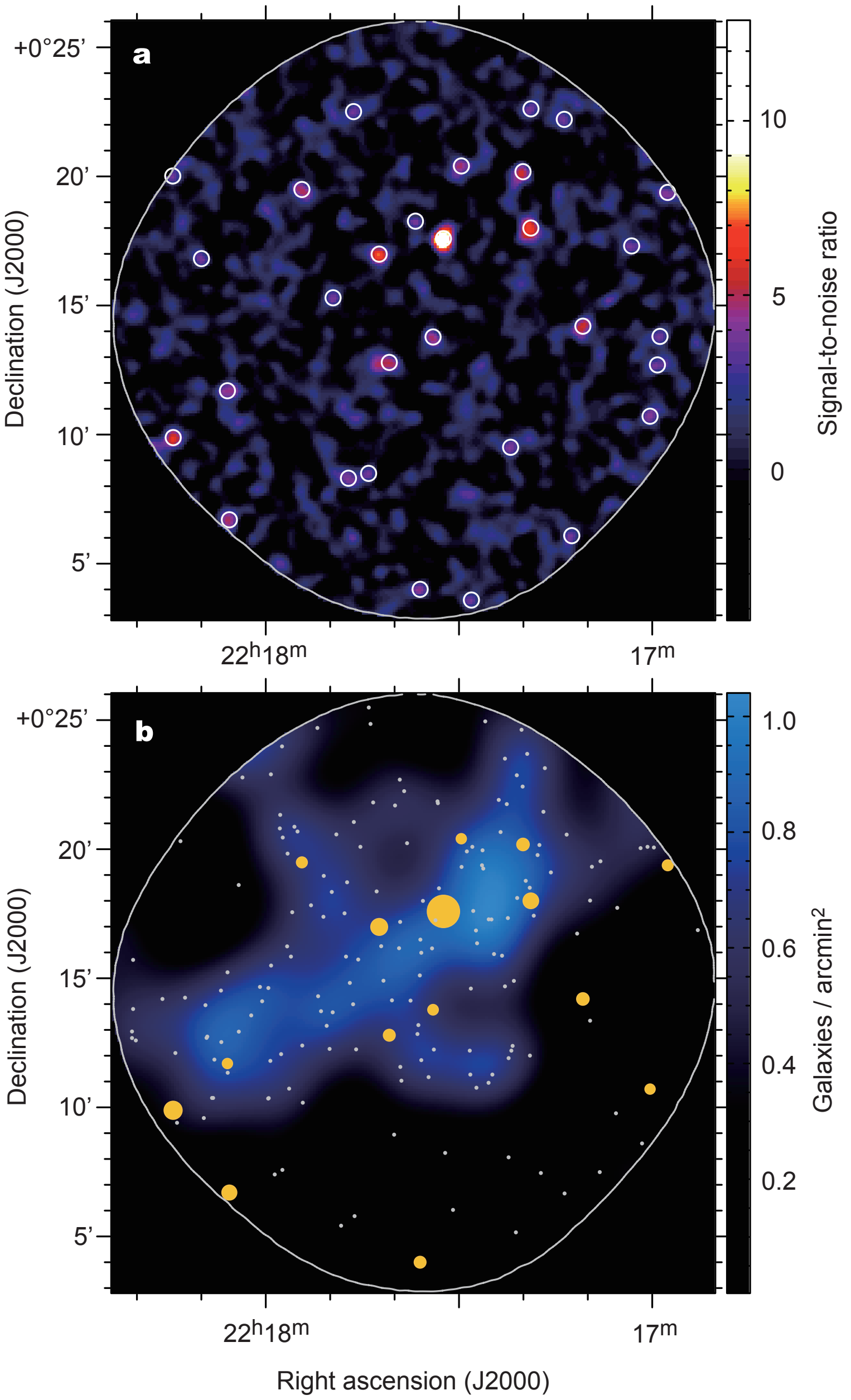

**Figure 1 The positions of 1,100-μm sources and Lyα emitters towards the SSA 22 protocluster region. a**, The colour scale shows the map of signal-to-noise ratio at 1,100 μm. The map shows 30 sources with signal-to-noise ratios ≥3.5 (circles). Observations of SSA 22 (field centre at RA = 22 h 17 min 36 s, dec. = +0° 15′ 00″ (J2000)) were obtained using the AzTEC camera[13], operating at 1,100 μm, mounted on the ASTE 10-m submillimetre telescope[14] during the July–September 2007 observing season. The data consist of a total of 42 h of integration time on source under excellent conditions (zenith atmospheric opacity at 220 GHz, $\tau_{220\ \mathrm{GHz}}$ = 0.01–0.10). This resulted

in a root-mean-square noise level of 0.68–0.99 mJy per beam over 390 arcmin$^2$. The point spread function of AzTEC on ASTE has a full-width at half-maximum of 28 ± 1 arcsec. **b**, The locations of the bright submillimetre galaxies with $S_{1,100\ \mu m} \geq 2.7$ mJy (orange filled circles) and the Lyα emitters at $z = 3.1$ (white dots). The sizes of the orange circles are proportional to their 1,100 μm fluxes. The number density field of the Lyα emitters is shown in the colour scale, highlighting the density enhancement of the Lyα emitters, which is thought to trace out the underlying large-scale structure at $z = 3.1$.

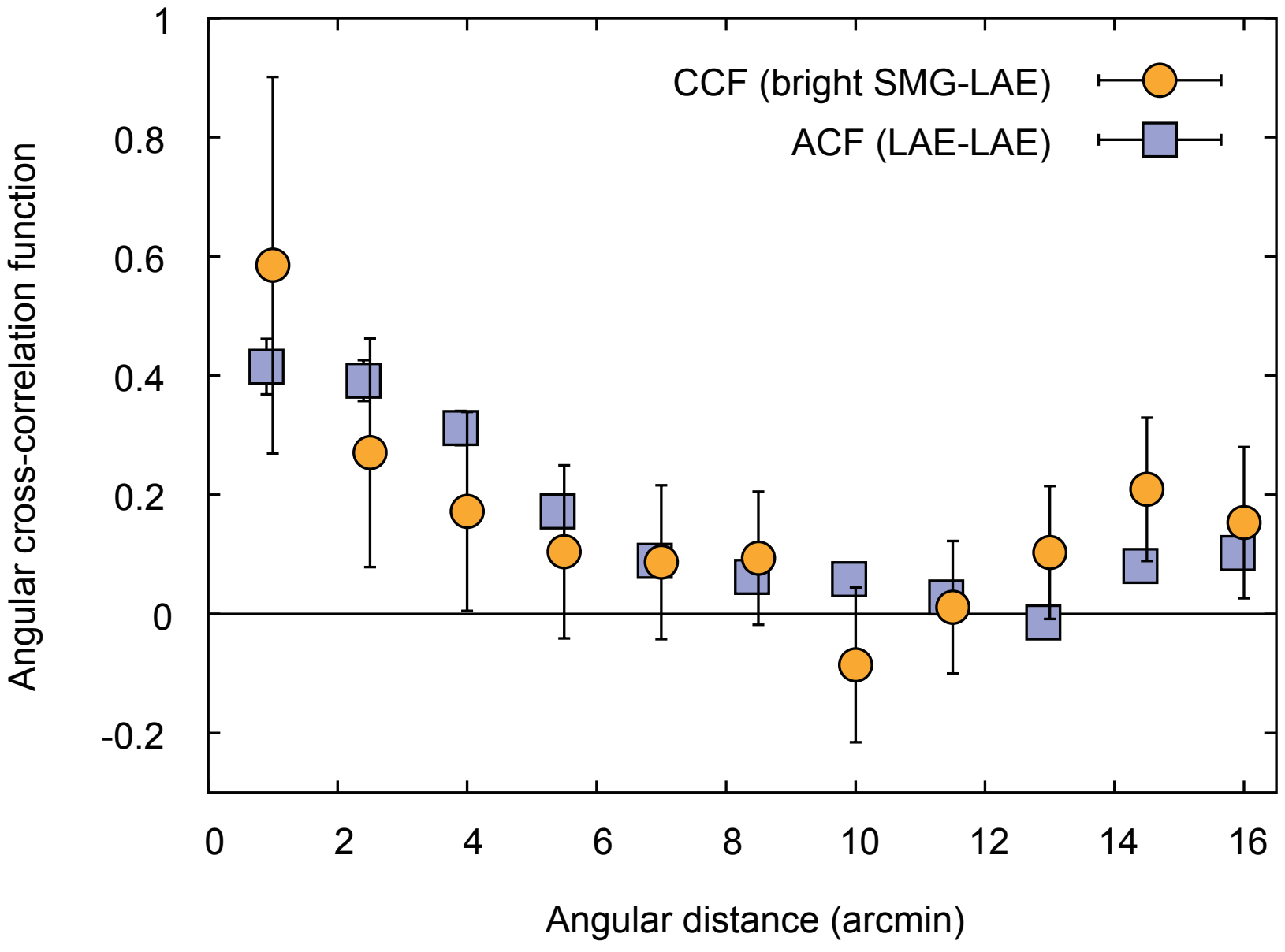

**Figure 2 Angular cross-correlation between submillimetre galaxies and Lyα emitters.** The two-point angular cross-correlation function shown here is computed for the 166 Lyα emitters and the 15 brightest ($S_{1,100\ \mu m} \geq 2.7$ mJy) submillimetre galaxies (orange circles). For reference, we also show the two-point angular autocorrelation function for the SSA 22 Ly-α emitters (blue squares). Small-number statistics prevent us from constraining the auto-correlation function well for the submillimetre galaxies. The correlation functions are computed using the estimator of ref. 29. The error bars are estimated from the root mean square of 1,000 bootstrap samples. See Supplementary Information for details.

# Supplementary Information

## 1. Calibration and analyses of the data obtained with AzTEC/ASTE.

Telescope pointing was checked every two hours using a nearby radio-loud quasar 3C446, resulting in a pointing accuracy better than 3 arcsec (ref. 30). Uranus and Neptune were observed two or three times per night for flux calibration, beam shape measurements, and array flat fielding. The point spread function of the AzTEC instrument on ASTE has a full-width at half-maximum of 28 ± 1 arcsec. Since emission from the atmosphere largely dominates the detector output, we removed the atmospheric noise fluctuations based on a principal component analysis technique (PCA cleaning)[19,29]. The response function of the PCA cleaning process to a point source is simulated, and then used for optimal filtering of the signal map for point-source detection in image processing. This resulted in an r.m.s. noise level of 0.68–0.99 mJy/beam over 390 arcmin$^2$. The fluxes of the detected sources were 'de-boosted' to correct for flux bias using the recipe described elsewhere[29,31]. Completeness in the survey was also computed by simulating the detection rate of 1,000 fake point sources per flux bin placed in the real cleaned signal map one by one. The completeness was found to be 50% at 2.7 mJy and 90% at 4.0 mJy (Supplementary Fig. 2S).

## 2. Angular cross-/auto-correlation functions.

The angular cross-correlation function (CCF) and auto-correlation function (ACF) shown in Fig. 2 are computed using the Landy & Szalay estimator[28]. For calculating the correlation functions, 25 sky realizations with 1,000 random sources are used. The error bars on the correlation functions are estimated from the r.m.s. among 1,000 bootstrap samples of the original catalogues. Integral constraint is taken into account for correcting the suppression of the amplitude of the correlation functions caused by poor sampling of large angular scales. Bins of 1.5 arcmin are used, but separations $\theta < 15$ arcsec are excluded to avoid the source confusion effect.

## 3. Significance level of the angular correlation between distributions of the SMGs and Lyα emitters.

We applied Monte Carlo simulations using the bi-dimensional Kolmogorov-Smirnov $D$-statistic[32] to assess the significance level of the apparent angular correlation between the distributions of the 15 bright SMGs and the 166 Lyα emitters. The $D$-statistic is used here to quantify the difference between the two distributions. We performed 10,000 trials of calculating the $D$-statistic for 15 random sources and the Lyα emitters, and compared these simulated $D$-statistics with that measured between the 15 bright SMGs and Lyα emitters ($D_{obs} = 0.222$). We found that 273 trials out of 10,000 have simulated $D$-statistics smaller than the observed $D$-statistic. This suggests that only 2.7% of the random distributions are better correlated to the Lyα emitter distribution than the real 15 SMGs (i.e., the significance level of the apparent correlation is 97.3% or 2.2σ).

## 4. Photometric redshift estimates.

There are 12 SMGs with 20 cm (Very Large Array) radio data available and 13 SMGs with 24 μm (Spitzer Space Telescope/MIPS) data available, and 5 SMGs of them have a counterpart both at 20 cm and 24μm. Positional uncertainties of the SMGs (error circles) are derived through Monte Carlo simulation to confine positions where the possible multi-wavelength counterpart(s) are likely located. We select a possible counterpart within the 2σ error circle of the AzTEC centroids (≈ 26″ diameter for 3.5 < $S/N$ < 4.5 and $\lesssim$ 20″ for $S/N$ > 4.5) and determine the 20-cm and 24-μm fluxes ($S_{20cm}$ and $S_{24\mu m}$) of the SMGs. We also give 2σ upper limits on $S_{20cm}$ and/or $S_{24\mu m}$ if no candidates are found. One should note that the fluxes at 20 cm and 24 μm are not always measured for the identical counterpart candidate because the 20-cm and 24-μm coverages are not well overlapped. This is why the redshifts shown in Supplementary Figs 4a and b are somewhat inconsistent. The fluxes at 20 cm and 24 μm dim rapidly as redshift increases while 1,100 μm fluxes are almost constant as a function of redshift[6] due to the difference in the local slopes of the spectral energy distribution (SED). For examining the redshift dependence of the fluxes, we used SEDs[33] modelled for 30 nearby (ultra-)luminous infrared galaxies ($L_{FIR}$ = $10^{11.2}$ – $10^{12.5}$ $L_{Sun}$).

---

# Supplementary Notes

## Additional references cited in Supplementary Information.

## Supplementary Figures

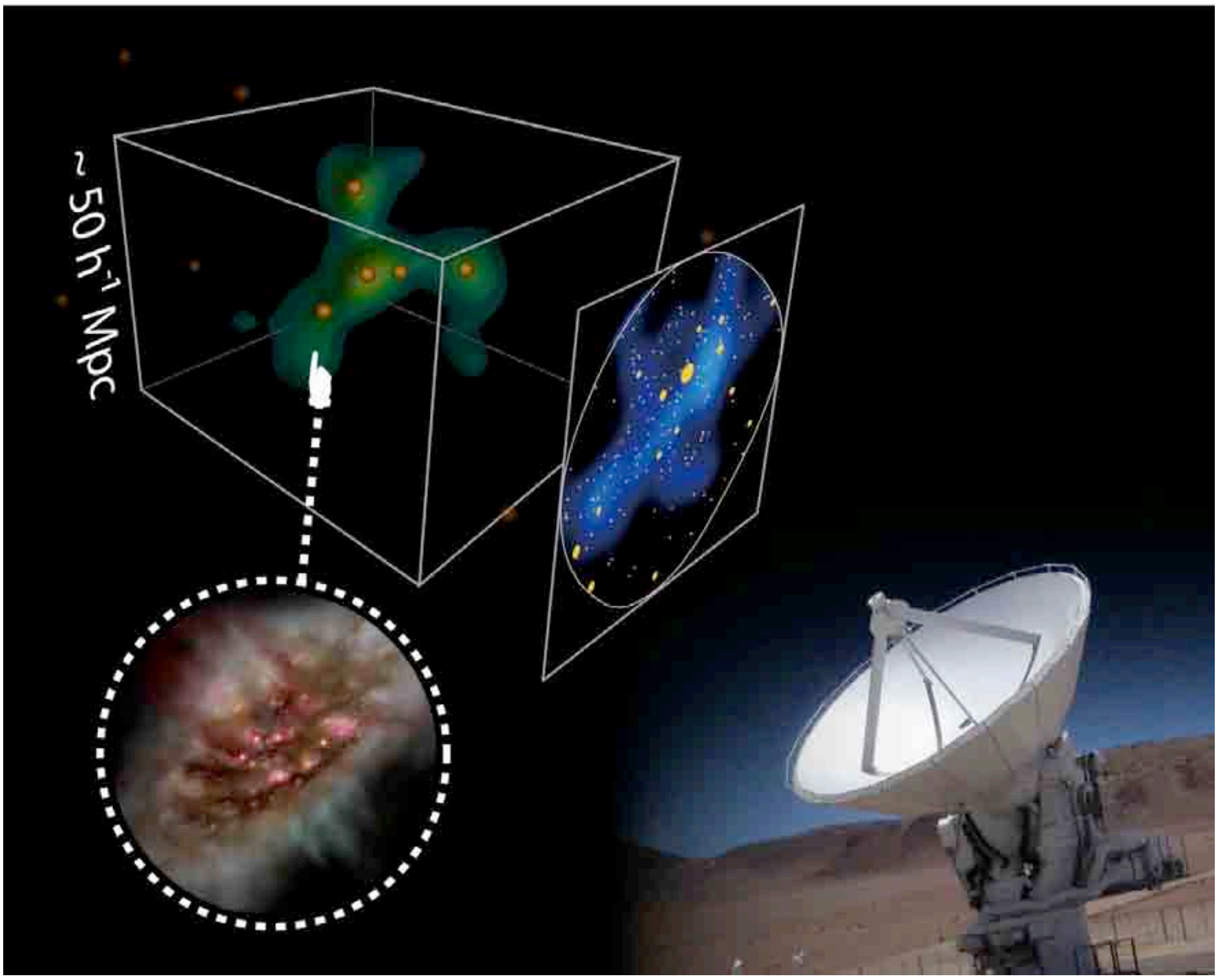

**Figure S1 | Schematic picture of this work.** The filamentary structure in green shown in the top-left corner represents the proto-cluster outlined by Lyman-α emitting galaxies in the SSA 22 field. We found an apparent clustering of submillimetre galaxies, which are believed to be massive dusty starburst galaxies (orange dots; an artist's conception of a submillimetre galaxy is shown in the bottom-left corner), towards the proto-cluster using the AzTEC camera mounted on the ASTE telescope (shown in the bottom-right corner). Although the 1,100-μm map shows only the projected distribution of the submillimetre galaxies on the plane of the sky, it is likely that some fraction of our submillimetre galaxies actually belongs to the proto-cluster, marking the local peak of underlying mass distribution.

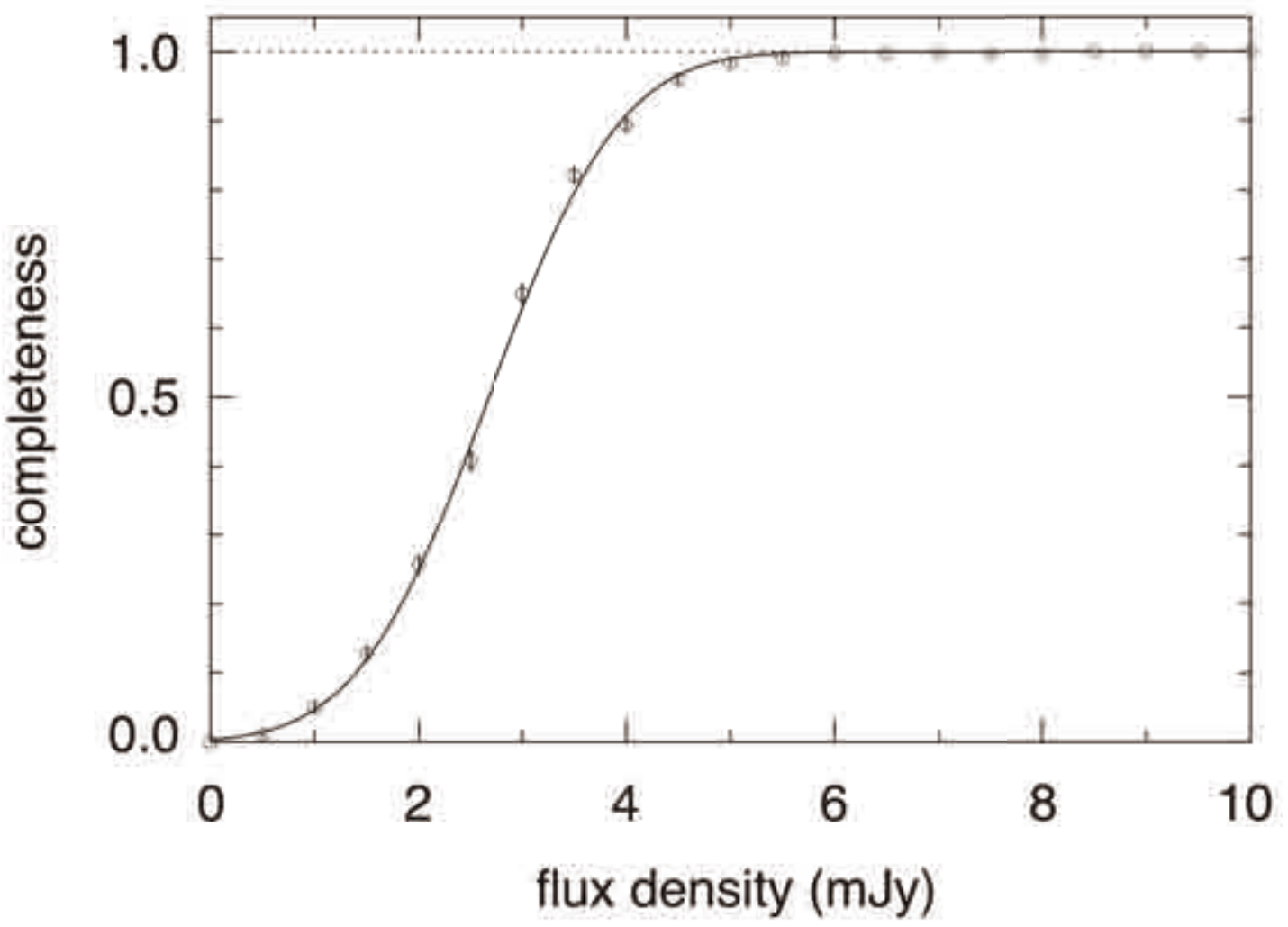

**Figure S2 | Completeness function.** The plots show the differential completeness derived through Monte Carlo simulations. The error bars on each plot are estimated from the binomial distribution. The solid curve is the best-fit curve given by (1/2) erf [ $(S - a) / b$ ], where $S$ is flux density measured in units of mJy, erf [ ] is the error function, $a = 2.67$ mJy, and $b = 1.41$ mJy.

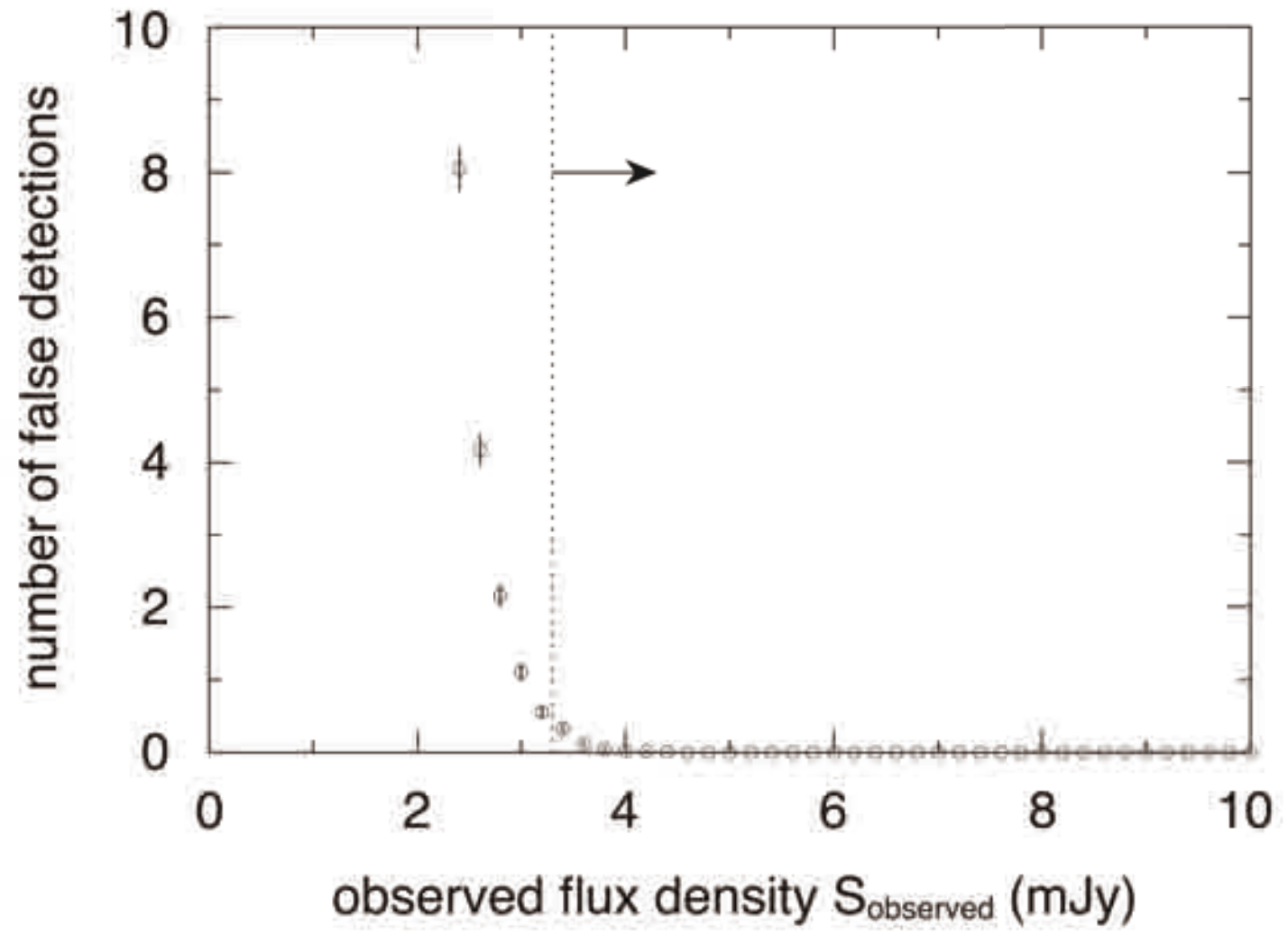

**Figure S3 | The number of false detections over the 390 arcmin$^2$ map.** The plots show the cumulative number of false detections above given flux density thresholds $S_{\mathrm{observed}}$ expected over the 1,100-μm map shown in Fig. 1. The numbers are estimated by extracting spurious sources from 100 simulated pure noise maps with the ordinary extracting algorithm. The details for simulation technique are described elsewhere[29]. The error bars are the 1σ Poisson uncertainties. The *observed* (before de-boosting) flux densities of the bright SMGs are 3.3 mJy or brighter (see Supplementary Table S1), and so it is likely that less than one fake source contaminates the bright sample (vertical dotted line and arrow).

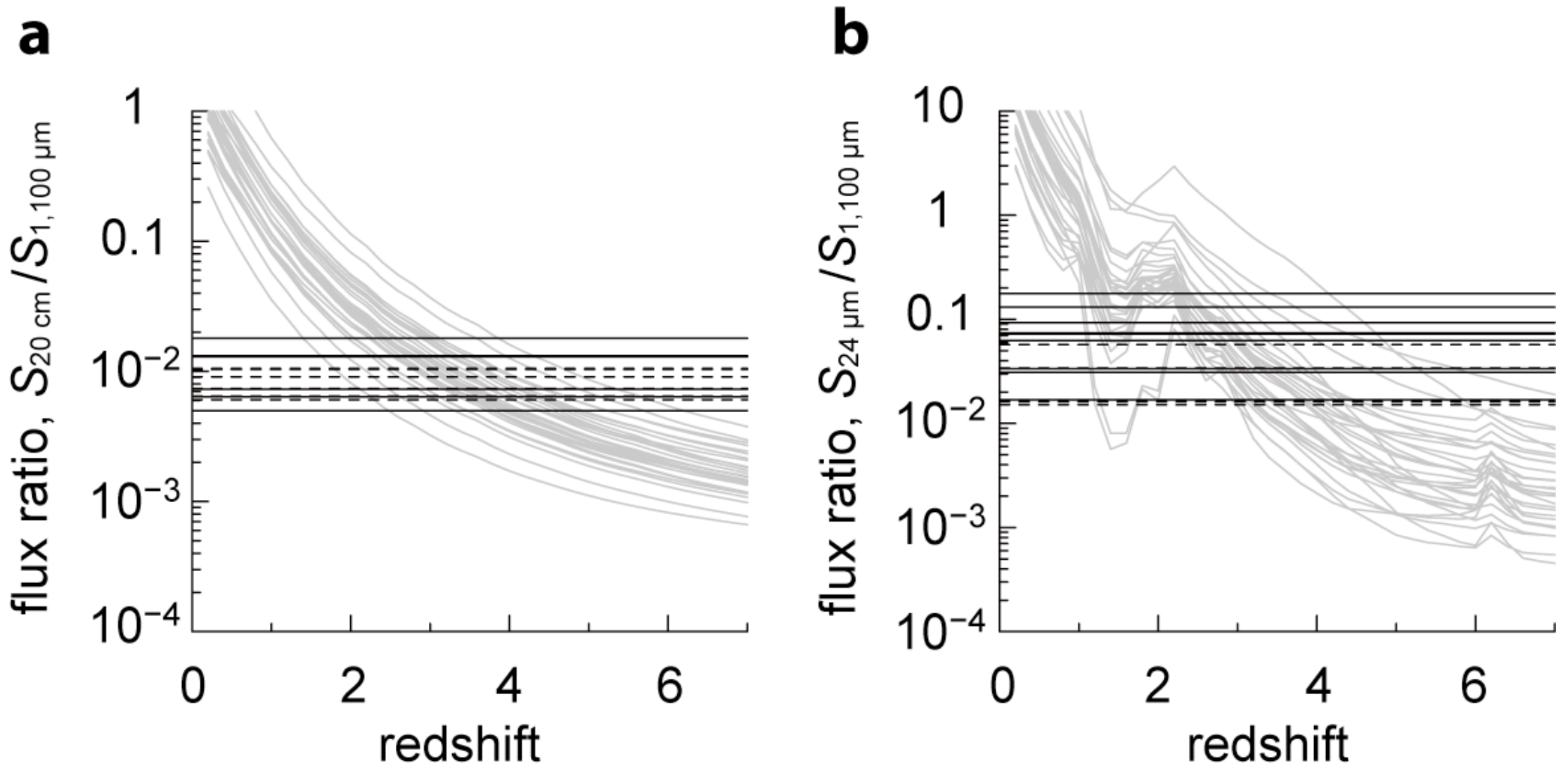

**Figure S4 | Photometric redshift estimates of the submillimetre galaxies towards SSA 22.** Horizontal lines show the flux ratios, (**a**) $S_{20cm}/S_{1,100\mu m}$ and (**b**) $S_{24\mu m}/S_{1,100\mu m}$, for each SMG. The 20 cm and 24 μm fluxes are obtained from the counterpart candidates within the 2σ error circle of the AzTEC centroids (≈ 26″ diameter for 3.5 < $S/N$ < 4.5 and ≲ 20″ for $S/N$ > 4.5) (solid lines), or from 2σ upper limits if no candidates are found (dashed lines). Solid curves in grey indicate the redshift dependence of the flux ratios predicted from spectral energy distributions[32] fit to 30 nearby infrared luminous star-forming galaxies. It is likely that some fraction of the SMGs are at $z = 3.1$ although there remain uncertainties in redshift space due to the scatter of the flux ratios predicted from the various templates of the spectral energy distribution.

## Supplementary Table

**Table S1 | Source positions and flux densities.**

| Source name | Coordinate (J2000) | | Flux density (mJy) | | S/N |
|---|---|---|---|---|---|
| | R.A. | Decl. | $S_{observed}$* | $S_{deboost}$† | |
| SSA22-AzTEC1 | 22:17:32.42 | +0:17:35.5 | 8.7 ± 0.7 | $8.4^{+0.8}_{-1.0}$ | 12.8 |
| SSA22-AzTEC2 | 22:17:42.38 | +0:16:59.3 | 4.9 ± 0.7 | $4.4^{+0.9}_{-0.8}$ | 7.2 |
| SSA22-AzTEC3 | 22:17:18.85 | +0:18:0.0 | 4.7 ± 0.7 | $4.1^{+1.0}_{-0.8}$ | 6.8 |
| SSA22-AzTEC4 | 22:18:14.37 | +0:9:53.1 | 5.6 ± 0.9 | $4.7^{+1.2}_{-1.0}$ | 6.2 |
| SSA22-AzTEC5 | 22:17:10.77 | +0:14:11.8 | 4.0 ± 0.7 | $3.3^{+1.0}_{-0.8}$ | 5.6 |
| SSA22-AzTEC6 | 22:17:20.07 | +0:20:11.0 | 4.0 ± 0.7 | $3.3^{+1.0}_{-0.8}$ | 5.6 |
| SSA22-AzTEC7 | 22:17:40.82 | +0:12:47.6 | 3.6 ± 0.7 | $3.1^{+0.8}_{-0.9}$ | 5.2 |
| SSA22-AzTEC8 | 22:18:5.65 | +0:6:42.0 | 4.9 ± 1.0 | $3.9^{+1.2}_{-1.2}$ | 5.0 |
| SSA22-AzTEC9 | 22:17:54.40 | +0:19:29.5 | 3.6 ± 0.7 | $2.9^{+0.9}_{-0.9}$ | 5.0 |
| SSA22-AzTEC10 | 22:17:34.03 | +0:13:46.8 | 3.4 ± 0.7 | $2.8^{+0.9}_{-0.9}$ | 4.8 |
| SSA22-AzTEC11 | 22:17:29.64 | +0:20:24.4 | 3.3 ± 0.7 | $2.7^{+0.9}_{-0.9}$ | 4.7 |
| SSA22-AzTEC12 | 22:17:36.04 | +0:4:0.2 | 4.0 ± 0.9 | $3.1^{+1.1}_{-1.1}$ | 4.5 |
| SSA22-AzTEC13 | 22:18:5.95 | +0:11:41.9 | 3.3 ± 0.7 | $2.7^{+0.9}_{-1.0}$ | 4.5 |
| SSA22-AzTEC14 | 22:17:0.34 | +0:10:42.6 | 3.7 ± 0.9 | $2.7^{+1.2}_{-1.2}$ | 4.2 |
| SSA22-AzTEC15 | 22:16:57.60 | +0:19:22.8 | 4.1 ± 1.0 | $2.9^{+1.4}_{-1.3}$ | 4.2 |
| SSA22-AzTEC16 | 22:17:21.98 | +0:9:30.5 | 2.8 ± 0.7 | $2.1^{+1.0}_{-0.9}$ | 4.0 |
| SSA22-AzTEC17 | 22:17:47.17 | +0:8:18.5 | 2.8 ± 0.7 | $2.1^{+1.0}_{-1.0}$ | 4.0 |
| SSA22-AzTEC18 | 22:17:49.56 | +0:15:18.0 | 2.6 ± 0.7 | $2.0^{+0.9}_{-1.0}$ | 3.9 |
| SSA22-AzTEC19 | 22:17:44.04 | +0:8:29.8 | 2.7 ± 0.7 | $2.0^{+1.0}_{-1.0}$ | 3.9 |
| SSA22-AzTEC20 | 22:17:36.76 | +0:18:16.0 | 2.6 ± 0.7 | $1.9^{+1.0}_{-1.0}$ | 3.8 |
| SSA22-AzTEC21 | 22:18:9.99 | +0:16:49.0 | 2.9 ± 0.8 | $2.1^{+1.2}_{-1.2}$ | 3.8 |
| SSA22-AzTEC22 | 22:17:3.20 | +0:17:18.5 | 3.0 ± 0.8 | $2.1^{+1.2}_{-1.2}$ | 3.8 |
| SSA22-AzTEC23 | 22:17:18.83 | +0:22:36.9 | 3.1 ± 0.8 | $2.2^{+1.3}_{-1.3}$ | 3.8 |
| SSA22-AzTEC24 | 22:18:14.44 | +0:20:1.0 | 3.7 ± 1.0 | $2.4^{+1.8}_{-1.9}$ | 3.7 |
| SSA22-AzTEC25 | 22:16:59.19 | +0:12:42.0 | 3.1 ± 0.8 | $2.1^{+1.4}_{-1.3}$ | 3.7 |
| SSA22-AzTEC26 | 22:17:12.49 | +0:6:5.1 | 3.5 ± 0.9 | $2.3^{+1.7}_{-1.7}$ | 3.7 |
| SSA22-AzTEC27 | 22:16:58.79 | +0:13:48.1 | 3.0 ± 0.8 | $2.1^{+1.3}_{-1.4}$ | 3.7 |
| SSA22-AzTEC28 | 22:17:46.37 | +0:22:30.6 | 2.8 ± 0.8 | $2.0^{+1.2}_{-1.3}$ | 3.7 |
| SSA22-AzTEC29 | 22:17:13.65 | +0:22:12.6 | 3.2 ± 0.9 | $2.1^{+1.5}_{-1.4}$ | 3.7 |
| SSA22-AzTEC30 | 22:17:28.07 | +0:3:35.5 | 3.5 ± 1.0 | $2.3^{+1.7}_{-1.9}$ | 3.6 |

We define AzTEC1-15 as the `bright' SMG sample with de-boosted flux densities equal to or greater than 2.7 mJy; whereas the remainder is the `faint' SMG sample. Note that the astrometric accuracy of the catalogue is ≈10 arcsec.

* Observed flux density before flux bias correction and the 1σ error.

† De-boosted flux density (corrected for the flux bias) and the 68% confidence interval.